\documentclass[prl,floatfix,twocolumn,showpacs,preprintnumbers,amsmath,amssymb,superscriptaddress]{revtex4}
\usepackage{color}

\usepackage{graphicx,psfrag} 
\usepackage{hyperref}
\usepackage{verbatim}
\usepackage[utf8]{inputenc}
\newcommand{\newc}{\newcommand}
\newc{\gsim}{\lower.7ex\hbox{$\;\stackrel{\textstyle>}{\sim}\;$}}
\newc{\lsim}{\lower.7ex\hbox{$\;\stackrel{\textstyle<}{\sim}\;$}}
\newc{\gev}{\,{\rm GeV}}
\newc{\mev}{\,{\rm MeV}}
\newc{\ev}{\,{\rm eV}}
\newc{\kev}{\,{\rm keV}}
\newc{\tev}{\,{\rm TeV}}

\newc{\mz}{M_Z}
\newc{\mpl}{M_*}
\newc{\mw}{m_{\rm weak}}
\newc{\nr}[1]{N^c_R{}_{#1}}

\newc{\qcd}{\,{\rm QCD}}

\newcommand{\eV}{{\, {\rm eV}}}
\newcommand{\keV}{{\, {\rm keV}}}

\newcommand{\GeV}{{\, {\rm GeV}}}

\definecolor{mypurple}{RGB}{164,64,214}

\usepackage{amsmath}
\def\beq{\begin{equation}}
\def\eeq{\end{equation}}
\def\bea{\begin{eqnarray}}
\def\eea{\end{eqnarray}}
\def\bitem{\begin{itemize}}
\def\eitem{\end{itemize}}
\newc{\ie}{{\it i.e.}}          \newc{\etal}{{\it et al.}}
\newc{\eg}{{\it e.g.}}          \newc{\etc}{{\it etc.}}
\newc{\cf}{{\it c.f.}}
\def\bar#1{\overline{#1}}

\def\inv{^{\raise.15ex\hbox{${\scriptscriptstyle -}$}\kern-.05em 1}}
\def\lbar{{\lower.35ex\hbox{$\mathchar'26$}\mkern-10mu\lambda}} 

\let\<=\langle
\let\>=\rangle

\let\+=\uparrow

\begin{document}   
\title{Twin Higgs Asymmetric Dark Matter} 
\author{Isabel Garc\'ia Garc\'ia}
\email{isabel.garciagarcia@physics.ox.ac.uk}
\affiliation{Rudolf Peierls Centre for Theoretical Physics, University of Oxford,
1 Keble Rd., Oxford OX1 3NP, UK}
\author{Robert Lasenby}
\email{robert.lasenby@physics.ox.ac.uk}
\affiliation{Rudolf Peierls Centre for Theoretical Physics, University of Oxford,
1 Keble Rd., Oxford OX1 3NP, UK}
\author{John March-Russell}  
\email{jmr@thphys.ox.ac.uk}  
\affiliation{Rudolf Peierls Centre for Theoretical Physics, University of Oxford,
1 Keble Rd., Oxford OX1 3NP, UK}
\affiliation{Department of Physics, Stanford University, Stanford, CA 94305, USA}

\begin{abstract}
We study Asymmetric Dark Matter (ADM) in the context of the minimal (Fraternal) Twin Higgs  
solution to the little hierarchy problem, with a twin sector with gauged $SU(3)' \times SU(2)'$, a twin
Higgs, and only third generation twin fermions.  Naturalness requires the QCD$^\prime$ scale
$\Lambda'_{\rm QCD} \simeq 0.5 - 20 \ {\rm GeV}$, and $t'$ to be heavy. 
We focus on the light $b'$ quark regime, $m_{b'} \lesssim \Lambda'_{\rm QCD}$, where QCD$^\prime$ is
characterised by a single scale $\Lambda'_{\rm QCD}$ with no light pions.  A twin baryon number
asymmetry leads to a successful DM candidate: the spin-3/2 twin baryon, $\Delta' \sim b'b'b'$, with
a dynamically determined mass ($\sim 5 \Lambda'_{\rm QCD}$) in the preferred range for the
DM-to-baryon ratio $\Omega_{\rm DM}/\Omega_{\rm baryon} \simeq 5$.
Gauging the $U(1)'$ group leads to twin atoms ($\Delta'$ - $\bar {\tau'}$ bound states) that
are successful ADM candidates in significant regions of parameter space, sometimes with observable changes
to DM halo properties. Direct detection signatures satisfy current bounds, 
at times modified by dark form factors.
\end{abstract}   
\maketitle 
\section{\label{intro}I. Introduction}
Despite overwhelming evidence for the existence of Dark Matter (DM), its precise nature remains a mystery.
Moreover, the closeness of DM and baryon energy densities, $\Omega_{\rm DM} \simeq 5 \Omega_{\rm baryon}$,  is fundamentally puzzling:  there seems to be no reason for these two quantities, a priori unrelated, to be so close to each other.
This puzzle motivates the idea of  Asymmetric Dark Matter (ADM)
\cite{Nussinov:1985xr,Gelmini:1986zz,Chivukula:1989qb,Barr:1990ca,Kaplan:1991ah,Hooper:2004dc,Kitano:2004sv,Cosme:2005sb,Farrar:2005zd,Suematsu:2005kp,Tytgat:2006wy,An:2009vq,Kaplan:2009ag},
based on the assumption that the present DM density is set by an asymmetry $\eta_{\rm DM}$ in the DM sector,
analogous to the baryon asymmetry $\eta_{\rm baryon}$.  Then
\begin{equation}
	\frac{\Omega_{\rm DM}}{\Omega_{\rm baryon}} = \frac{m_{\rm DM}}{m_{N}} \frac{\eta_{\rm DM}}{\eta_{\rm baryon}}
\label{eq:ratioOmega}
\end{equation}
where $m_N$ is the nucleon mass and $m_{\rm DM}$ the mass of the DM particle.
A linked asymmetry of the same order in both sectors, $|\eta_{\rm DM}| \sim |\eta_{\rm baryon}|$, is relatively
easy to achieve, but a successful explanation of $\Omega_{\rm DM} / \Omega_{\rm baryon}$ requires a \emph{reason}
for $m_{\rm DM} \sim m_{N}$, and this has been the Achilles heel of ADM model building.

Another pressing worry is the LHC naturalness problem:  why have the new particles and/or dynamics that
stabilise the weak scale not been observed?  The Twin Higgs (TH) solution to this \emph{little} hierarchy problem
is based on the realisation of the Higgs boson as a pseudo-Nambu-Goldstone of an approximate global $SU(4)$ symmetry \cite{TwinHiggs}\cite{Chacko2005,ChackoNomura,Barbieri2005,Craig:2013fga}.
The TH mechanism introduces a Standard Model (SM) \emph{neutral} sector, the \emph{twin} sector, that is an approximate copy of the SM, and the Higgs sector of the theory must respect, at tree-level, an $SU(4)$ global symmetry that acts on the two (visible and twin sector) Higgs doublets.  A $\mathbb{Z}_2$ between sectors imposes all couplings to be equal and ensures that radiative corrections to the Higgs soft mass squared are $SU(4)$ symmetric.
The global $SU(4)$ is only broken at 1-loop and the $\mathbb{Z}_2$ must be broken, explicitly or otherwise, for the vacuum expectation value (vev) of the twin and visible sector Higgs (denoted $f$ and $v \approx 246 \GeV$ respectively) to be different.
As the possibility of $f=v$  is ruled out by Higgs coupling measurements,
the minimal fine-tuning in the electroweak sector is then given by $\sim \frac{2 v^2}{f^2}$
(only $\sim 20\%$ for $f/v \approx 3$, the minimum experimentally allowed ratio).

The TH mechanism does not require the twin sector  to be an exact copy of the SM.  A
minimal realisation, the Fraternal Twin Higgs (FTH)\cite{FraternalTH}, only requires
in the twin sector $SU(3)'\times SU(2)'$ interactions, top and bottom quarks ($Q'$, $t'_R$, $b'_R$),
and lepton ($L'$) and Higgs ($H'$) doublets.
Twin right-handed leptons are not required but may be added, and
a $U(1)'$ gauge group is \emph{not} required by naturalness, although it remains an accidental global symmetry.
Masses of twin fermions are set by their Yukawa couplings and the ratio $f/v$.
Naturalness requires a twin top Yukawa $y_{t'} \simeq  y_{t}$,
but only imposes $y_{i'} \ll 1$ ($i' \neq t'$).
Most important for us, values of the $g'_3$ gauge coupling consistent with naturalness imply
a QCD$^\prime$ scale $\Lambda^\prime_{\rm QCD} \sim 0.5 - 20 \GeV$ for a
$5 \tev$ cutoff \cite{FraternalTH}.  (The theory needs UV-completion at some scale $M_{\rm UV} \lesssim 4 \pi f$.)

The purpose of this letter is to explore the possibility of ADM in the FTH context~\footnote{Related investigations of DM in Twin Higgs models have been carried out by other groups \cite{Craig:2015xla}\cite{Farina:2015uea}.}.
We work in the regime, $m_{b '} \lesssim \Lambda^\prime_{\rm QCD}$, where the twin QCD$^\prime$
theory is determined by a single scale, arguing that the baryon $\Delta ' \sim  b^\prime b^\prime b^\prime$, either on its
own, or in an atomic bound state with a $\bar{\tau^{\prime}}$ in the gauged $U(1)'$ case, is a successful ADM candidate.

\section{II. Stable \& Relativistic Twins}

Within the FTH scenario, the twin sector respects three accidental
global symmetries: twin baryon number ${\rm B'}$, lepton number
${\rm L'}$ and `charge' ${\rm Q'}$. If these are not too badly
broken by higher dimensional operators (HDOs), as we will assume,
then the lightest twin particles carrying these quantum numbers will be
cosmologically stable states. Twin $CP$ could be a good discrete symmetry
of the twin sector, although both
$P$ and $C$ are violated by $SU(2)'$ interactions.

We consider massive $\tau '$ but allow for heavy or massless $\nu '$,
usually with $m_{\tau'} + m_{\nu'} < m_{W'}$ so that $W'^\pm$ gauge bosons decay,
although a possibly interesting scenario arises if $m_{\tau'} + m_{\nu'} > m_{W'}$ and
$W'^\pm$ are stable. For $m_{b '} \lesssim
\Lambda^\prime_{\rm QCD}$, the lowest QCD$^\prime$ states are
$\bar {b '} b '$ mesons, the lightest being a pseudoscalar $\hat
\eta$ and a scalar $\hat \chi$ with masses $m_{\hat \eta} \approx (2
- 3) \Lambda^\prime_{\rm QCD}$ and $m_{\hat \chi} \approx 1.5 m_{\hat
\eta}$ \cite{1flavorQCDhadrons}.   (A distinctive feature is the absence of
pseudo-Nambu-Goldstone bosons due to the chiral anomaly.)
The glueball spectrum is heavier and only weakly mixed
with the mesons, with the lightest being a $0^{++}$ state of mass $m_0 \sim 7
\Lambda^\prime_{\rm QCD}$ \cite{Morningstar,Chen}.
Meson/glueball
states decay quickly via $SU(2)^\prime$ interactions to $\bar{\nu '} \nu'$ pairs if $m_{\nu '} \approx 0$
(and multi-$\gamma'$ states if $U(1)'$ is gauged)
and lighter mesons/glueballs,
or to SM states via twin-scalar$-$Higgs mixing \cite{Juknevich,FraternalTH}. Independently
of $m_{\nu '}$, the lightest twin meson $\hat \eta$ may
decay very fast via dimension-six HDOs that preserve total $CP$, of the
form $\sim (\bar q \gamma^5 q \ \bar {b'} \gamma^5 b')  /M^2$,
where $q$ denotes SM quarks (for $M \sim  10 \tev$,
this gives a lifetime $\tau_{\hat \eta}^{-1} \sim 10^{-14} \ {\rm s}$).

The spin-3/2 twin $\Delta '$ baryon with mass
$m_{\Delta '} \approx 5 \Lambda^\prime_{\rm QCD}$ \cite{1flavorQCDhadrons} and $U(1)'$
global charge $-1$, is naturally extremely long-lived since it is the lightest ${\rm B'} \neq 0$ object.
Moreover, the leading HDO violating SM and twin baryon numbers but preserving a linear combination is dimension-12,
resulting in a lifetime
$\tau_{\Delta'} \sim 10^{26} \ {\rm s}$ for $m_{\Delta '} \sim 10 \GeV$ and $M \sim 10 \tev$.
Thus even in the presence of HDOs, $\Delta'$ can be stable on cosmological timescales.
For the purposes of this paper we assume that the $\Delta'$ is the only twin baryon number carrying state
with a cosmologically relevant lifetime.  (The presence of heavier stable twin baryon states would not
qualitatively change our conclusions.)

Dark radiation (DR) contributions to the number of effective neutrino species, $\Delta N_{\rm eff}$,
can arise from light twin neutrinos, and twin photons when $U(1)'$ is gauged.
Due to the extremely fast decay of the lightest twin meson $\hat \eta$ into SM states naturally present via HDOs, we expect the $\nu'$ and $\gamma'$ sectors to remain in equilibrium with the SM after the QCD$^\prime$ phase transition, even for values of $\Lambda_{\rm QCD}^\prime$ as small as $\sim 0.5 \GeV$.
As a result, in the case of $m_{\nu '} \approx 0$ and \emph{no} gauged $U(1)'$ we expect a
contribution to $\Delta N_{\rm eff}$ of $\approx 0.075$ (as argued in Section VIII of \cite{THSymDM})
and of $\approx 0.16$ when twin photons are also present.
Notice these are the minimum possible contributions to $\Delta N_{\rm eff}$ and are compatible with the current measured value
$\Delta N_{\rm eff} - \Delta N_{\rm eff, SM} \simeq 0.1 \pm 0.2$ \cite{Ade:2015xua}, although future experiments may achieve an accuracy of $\sim 0.05$ \cite{Hannestad:2014voa,Bashinsky:2003tk} and therefore probe these two scenarios.

\section{III. Twin Baryon \&  $W^\prime$ Dark Matter}

The ADM scenario necessarily has a linked asymmetry in SM- and twin-sector quantum numbers.  The generation
of such an asymmetry is a UV issue --- here we simply assume that it is present. 
In addition, ADM requires efficient annihilation of the symmetric component of stable DM states, so that
the final DM abundance is dominantly set by the asymmetry.
In our case, annihilation of the symmetric component of the twin baryon states happens efficiently via twin
strong interactions.  Sufficiently heavy $\tau '$ and $\nu'$
species also annihilate efficiently, mainly to $\bar {b '} b '$ states
(see Figure 2 in \cite{THSymDM}).  The ${\rm QCD}'$ phase transition for $m_{b'} \lesssim \Lambda^\prime_{\rm QCD}$ is a
smooth crossover~\cite{Pisarski:1983ms,Alexandrou:1998wv,Fromm:2011qi},
so we expect neither significant non-equilibrium dynamics
nor entropy production affecting relic densities.

A twin baryon number asymmetry implies
an asymmetric relic population of $\Delta'$ baryons.
If $\eta_{\rm Q'} = 0$, then the (ungauged) charge
density of the $\Delta'$ population must be balanced by a population of
twin charged states. So, if the $\Delta'$ baryons are to be the only significant
DM component, either $m_{\tau '}\approx 0$ so
that an asymmetric abundance of these can exist as DR, 
or we must have a compensating asymmetry in (global) twin charge, 
$\eta_{\rm Q'} \simeq - \eta_{\rm B'}$. 
Depending on UV dynamics there may be a non-zero twin lepton asymmetry
setting an asymmetric $\nu'$ DR relic density (the $\tau'$
density is fixed by $\eta_{\rm B'}$ and $\eta_{\rm Q'}$).

As anticipated in Section II, $m_{\Delta '} \approx 5 \Lambda^\prime_{\rm QCD}$ \cite{1flavorQCDhadrons},
which translates into
$\eta_{{\rm B'}} / \eta_{\rm baryon} \approx m_N / \Lambda'_{\rm QCD}$,
with $\Lambda'_{\rm QCD} = 0.5 - 20 \GeV$ \cite{FraternalTH}.
Thus this framework allows for a successful realisation of ADM in which the mass of the DM particle
is not \emph{tuned} to be $\mathcal{O}(10 \GeV)$, but rather is set by the confinement scale of the DM sector, whose
range is restricted directly by naturalness arguments.
The value of $y_{b '}$ is irrelevant for the DM mass as
long as $m_{b '} \lesssim \Lambda^\prime_{\rm QCD}$ is realised.
DM in this framework is then made of \emph{individual} $\Delta '$ baryons.
Bound states, if they exist in the spectrum, will not form in the early
universe, since the only states parametrically lighter that could be
emitted in the binding processes are $\nu '$ or light SM states, but these both
only interact via tiny sub-weak interactions.
Moreover, we find that even in the presence of twin photons, radiative capture is not fast enough to give a 
significant population of ${\Delta '}-{\Delta '}$ bound states as the electric and magnetic dipole radiative capture rates vanish.
(This situation can be significantly different when lighter generations are present, in which case bound states may form
allowing for a scenario of nuclear DM \cite{HardyNDM,KrnNDM}.)

Regarding $\Delta '$ self-interaction bounds we have, parametrically,
$\sigma_{\Delta '}/m_{\Delta '} \sim (\Lambda'_{\rm QCD})^{-3} \sim
10^{-3} -10^{-8} \ {\rm cm^2 \ g^{-1}}$
for $\Lambda^\prime_{\rm QCD} = 0.5 - 20 \GeV$, well below the current
experimental upper bound of $\sim 0.5 \ {\rm cm^2 \ g^{-1}}$ \cite{DMselfint}.

Finally, in the case where $m_{\tau '} + m_{\nu '} > m_{W'}$, ${W'}^\pm$
are also a stable states, and even if $\eta_{{\rm B'}} = -
\eta_{{\rm Q'}}$, an asymmetric population of $\tau '$ ($\bar {\tau '}$)
states could survive, whose charge is balanced by an equal number of
asymmetric ${W'}^+$ (${W'}^-$) states. Notice that for small values
$f/v \approx 3 - 5$ (see Figure 4 in \cite{THSymDM}), annihilation of
the symmetric populations of $\tau '$, $\nu '$ and ${W'}^\pm$ occurs
very efficiently. For this latter possibility to be realised without
introducing significant extra tuning, one would need $m_{\tau '},
m_{\nu '} \sim 10^2 \GeV$ (since $m_{W'}\approx (f/v) m_{W}$), above the
mass range where ADM scenarios work most naturally.  
(Scattering cross sections of such states off SM nucleons
via the Higgs portal are $\lsim 10^{-45} \ {\rm cm^2}$ for
$f/v \gsim 4$, close to current bounds.)

\subsection{III.2. Direct Detection}

Scattering of $\Delta '$ baryons off SM nucleons happens via Higgs exchange or by exchanging a twin scalar state ($\hat \chi$ meson or
$0^{++}$ glueball) that mixes with the Higgs.  Couplings between scalar mesons/glueballs and a pair of twin baryons are unknown and require dedicated lattice computation.  We find that within a reasonable range for the couplings and mixing angles either Higgs exchange
or meson/glueball exchange can dominate the scattering.  We therefore separately consider the processes (ignoring interference effects)
to give an idea of the possible scattering cross sections.

In the case where Higgs exchange dominates, the spin independent scattering cross section is given by
\begin{equation}
	\sigma_h^{\rm SI} \approx \frac{1}{\pi}
	\mu_{N \Delta '}^2 \frac{(f_N m_N)^2}{m_h^4 v^4} \frac{(m_{\Delta '} f_{\Delta '})^2}{(f/v)^4}
\label{eq:sigmahiggs}
\end{equation}
where $\mu_{N \Delta '} = m_N m_{\Delta '} / (m_N + m_{\Delta '})$ is the reduced mass of the $\Delta '$-nucleon system.
$f_N \approx 0.32$ \cite{fNWagner, fNlattice, Crivellin:2013ipa}
and $f_{\Delta '} = (2 + 87 f_{b '})/31$ (following \cite{SVZ}) are the effective Higgs  couplings to nucleons and $\Delta '$ baryons,
respectively, where $f_{b '}$ is the dimensionless part of the matrix element
of $b '$ in $\Delta '$.
In the light $b '$ case, one expects $f_{b '} \ll 1$ albeit its exact value requires dedicated lattice study.
In the case where the dominant process is meson exchange, the cross section can be written as
\begin{equation}
	\sigma_{\hat \chi}^{\rm SI} \approx \frac{1}{\pi} \mu_{N \Delta '}^2 \frac{(f_N m_N)^2}{m_{\hat \chi}^4 v^2} {\lambda^\prime}^2 {\theta '}^2
\label{eq:sigmameson}
\end{equation}
where $\lambda^\prime$ is the coupling between $\hat \chi$ and a pair of $\Delta '$ baryons and
$\theta '$ is the Higgs-$\hat \chi$ mixing angle
\begin{equation}
	\theta ' = \frac{f_{\hat \chi} m_{\hat \chi}}{2 f (f/v)} \frac{{\mathcal{F}}_{\hat \chi}}{m_h^2 - m_{\hat \chi}^2}
\end{equation}
with ${\mathcal{F}}_{\hat \chi}$ the $0^{++}$ meson decay constant that we define as
${\mathcal{F}}_{\hat \chi} \equiv a' m_{\hat \chi}^2$
(with $a '$ an unknown dimensionless constant)
and $f_{\hat \chi} = (2 + 58 \tilde{f}_{b '})/31$ accounts for the effective coupling between meson and Higgs. 
Numerical evaluation shows that for $\lambda^\prime \lesssim 1$ Higgs exchange dominates,
whereas for $\lambda^\prime \gtrsim 4 \pi$ (the NDA value) meson exchange provides the leading interaction.
In the event of glueball exchange being the dominant process, the scattering cross
section is given by Eq(\ref{eq:sigmameson}) after performing the appropriate substitutions.

\begin{figure}[h!]
  \includegraphics[scale=0.52]{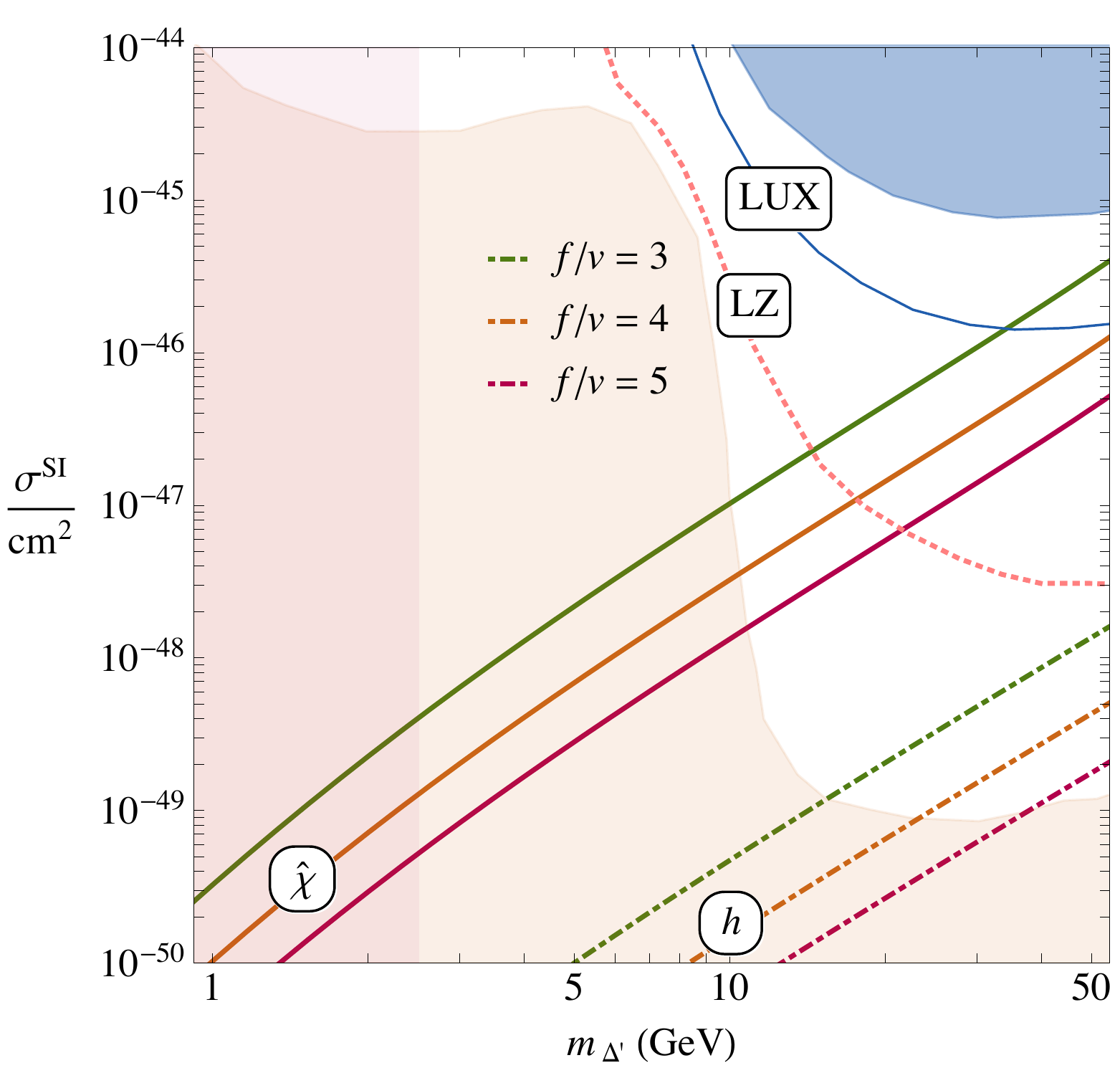}
\caption{\label{fig:sigma}
	Illustrative range of possible spin independent scattering cross sections of $\Delta '$ baryons off SM nucleons when either Higgs
	or $\hat \chi$ meson exchange dominates (dashed and thick lines respectively). 
	We take $m_{\hat \chi} = 3 \Lambda'_{\rm QCD}$, $\lambda ' = 4 \pi$, $a'=1$,  $f_{b '} = 0$
	and $\tilde f_{b '} = 0.1$ for illustration.
	Blue: LUX excluded \cite{LUXresults}; blue line: LUX projected sensitivity (300 live-days) \cite{LUXprojection};
	orange: neutrino background \cite{snowmass2013}; pink dotted line: LZ sensitivity \cite{snowmass2013};
	pink: values of $m_{\Delta '}$ (equivalently, of $\Lambda^\prime_{\rm QCD}$) that imply extra tuning \cite{FraternalTH}.
	} 
\end{figure}

Figure~\ref{fig:sigma} shows these spin independent scattering cross sections for particular choices of the unknown
parameters.  To illustrate the range possible we have chosen the minimum Higgs-exchange cross section (i.e. $f_{b '}=0$), while
for meson exchange we have selected reasonably large values of the parameters.  Note that different choices allow Higgs or glueball exchange to dominate. 
A significant portion of parameter space is covered by the neutrino floor, in particular the region $m_{\Delta '} \approx 5 \GeV$
that would allow for $\eta_{{\rm B'}} \approx \eta_{\rm baryon}$.
For values $m_{\Delta '} \approx 10 - 50 \GeV$, that correspond to $\eta_{{\rm B'}} / \eta_{\rm baryon} \approx 0.5 - 0.1$,
predicted cross sections escape the neutrino background and sit close to (or within)
the region that will be probed by next-generation experiments such as LZ \cite{snowmass2013}.

\section{IV. Twin Atoms}

\begin{figure}[h!]
  \includegraphics[scale=0.48]{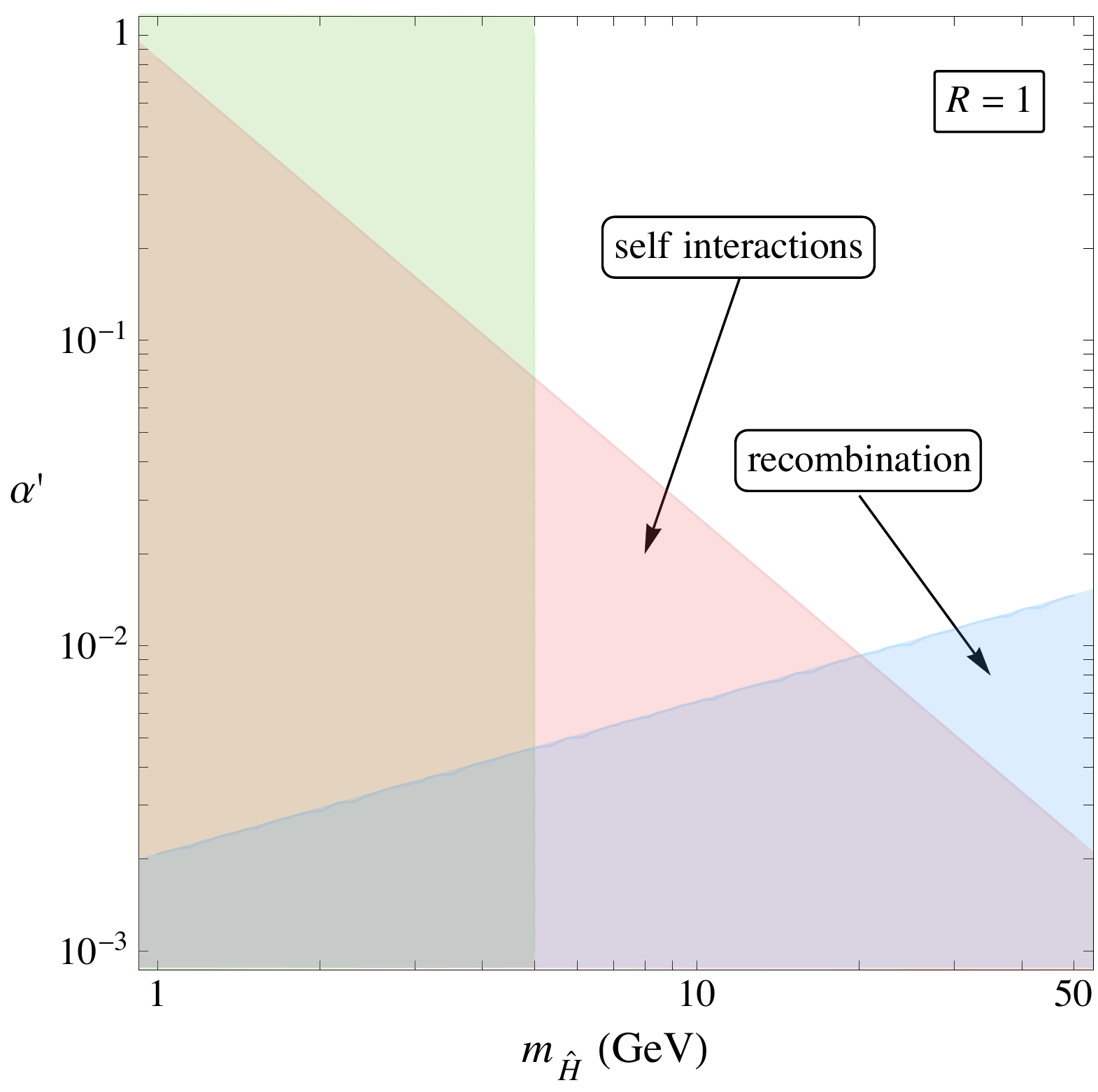}
\caption{\label{fig:atomR1}
Constraints in the $\alpha', m_{\hat H}$ plane, for a ratio $R = m_{\Delta '} / m_{\tau '} =1$.
		Blue: twin recombination is inefficient, an ionised fraction
		$\gtrsim 0.1$ remaining;
		pink: self-interaction cross section is $\gtrsim 0.5 \ {\rm cm^2 \ g^{-1}}$;
		green: twin atom masses small enough that significant extra tuning is present.} 
\end{figure}

\begin{figure}[h!]
  \includegraphics[scale=0.48]{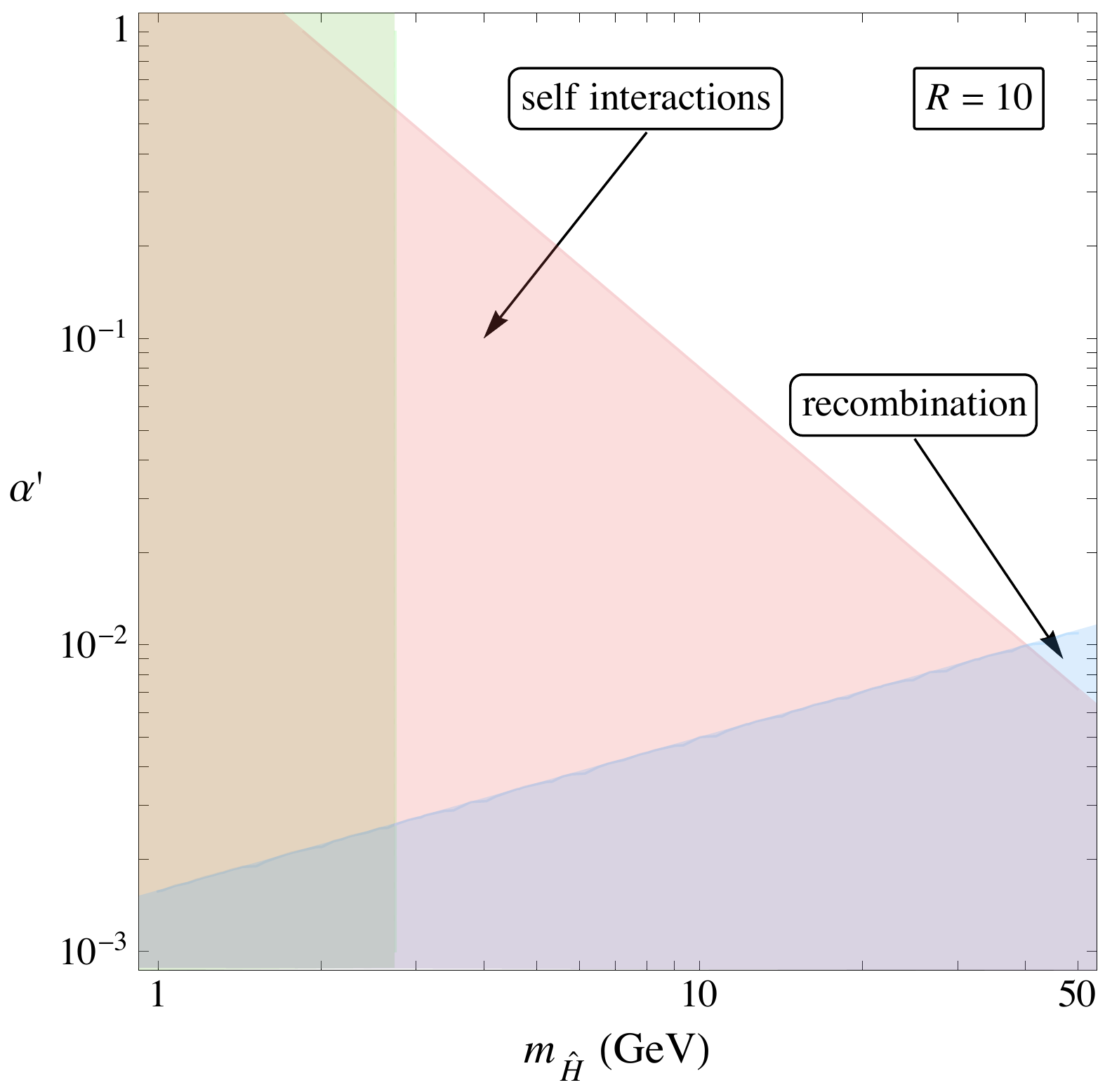}
\caption{\label{fig:atomR10}
As Figure~\ref{fig:atomR1} but for a ratio $R = m_{\Delta '} / m_{\tau '} =10$.} 
\end{figure}

Once the $U(1)'$ group is gauged, the physics becomes substantially
richer. Twin-charge neutrality of the Universe requires $\eta_{{\rm Q'}}
= 0$, which means that a ${\rm B'}$ asymmetry resulting in a non-zero
asymmetric population of $\Delta '$ baryons must be balanced by an
${\rm L'}$ asymmetry, such that an equal asymmetric population of $\bar
{\tau '}$ is present (we here assume that $W^{\prime \pm}$ are unstable).  Due to twin electromagnetic interactions, the
asymmetric populations of $\Delta '$ and $\bar {\tau '}$ states may form bound states.
In fact, the late-time DM population \emph{must} consist of
overall-neutral `twin atoms', rather than a plasma of charged states,
for values of the twin electromagnetic coupling $\alpha '$ that are not extremely small;
otherwise, the long-range interactions between DM particles result in
plasma instabilities that strongly affect Bullet Cluster-like
collisions~\cite{Heikinheimo:2015kra,Gabrielli:2015hua,Feldstein}. Requiring that efficient
twin recombination takes place imposes
non-trivial constraints on the sizes of
$\alpha '$ and the mass of the twin atom $\hat H$~\cite{CyrRacine:2012fz}.
Further constraints are present due to DM self-interactions:  Low energy 
atom-atom scattering processes have cross sections
$\sigma \approx 10^2 (a'_0)^2$ where
$a'_0 = (\alpha ' \mu_{\hat H})^{-1}$ is the atomic Bohr radius
and $\mu_{\hat H}$ the reduced mass of the atomic system,
although the exact value of $\sigma$ depends strongly on the ratio 
$R \equiv m_{\Delta '} / m_{\tau '}$ for values $R \gtrsim 15$ \cite{Cline2013}.
We impose the constraint $\sigma / m_{\hat H} \lesssim
0.5 \ {\rm cm^2 \ g^{-1}}$~\cite{DMselfint} applicable to
contact-like DM scattering, since the effect of hard scatterings generally
dominates over soft or dissipative processes for atom-atom scattering
in the regimes we consider.  Figures~\ref{fig:atomR1} and
\ref{fig:atomR10} show constraints from recombination~\cite{CyrRacine:2012fz} and DM self-interactions, for
ratios $R \equiv m_{\Delta '} / m_{\tau '} = 1$ and $10$ respectively.


For values of $\alpha '$ and $m_{\hat H}$ satisfying recombination and
self-interaction constraints, and for the parameter ranges we consider, annihilation
of the symmetric populations of
$\Delta '$ and $\tau '$ happens very efficiently. As can be seen from
Figures~\ref{fig:atomR1} and \ref{fig:atomR10}, the minimum value
of $\alpha '$ consistent with all constraints is $\alpha ' \approx
10^{-2}$, in which case the twin atom mass is constrained to be $m_{\hat
H} \approx 20, 40 \GeV$ for $R=1,10$ respectively.  This results in
binding energies of order $\mathcal{O}(10^2) \ \kev$, and a hyperfine
splitting of the first atomic energy level of order $\Delta E \sim 10
\eV$.

Before twin sector recombination occurs, the $\Delta'$ and $\bar{\tau'}$
are coupled to the twin photon bath, constituting a dark plasma that
can undergo `dark acoustic oscillations'~\cite{CyrRacine:2012fz}. If
twin sector recombination is late enough, these oscillations can leave
an imprint in the power spectrum of baryonic matter. However, since
$\alpha' \gtrsim 10^{-2}$ in our allowed regions, the binding energy of
our twin atoms is sufficiently high ($\gg 10 \keV$) that twin recombination is always
too early to realise this possibility.

Another possibility is that, after dark recombination, molecular bound states may form at
lower temperatures. However, radiative capture of two neutral atoms to a `dark hydrogen molecule'
is very suppressed~\cite{Hirata:2006bt}, with molecule formation requiring that there is an
abundance of charged particles to catalyse the reactions.  Given the constraints
that must already be satisfied, our estimates indicate
that a significant proportion of molecules will not be formed, either in the
early universe, or in halos.

We remark that most of the physics discussed in this section is not specific to FTH models,
relying only on asymmetric DM charged under a dark $U(1)$.
There is a large body of literature on the physics of such `dark atoms',
e.g.~\cite{Goldberg:1986nk,Feng:2009mn,Kaplan:2009de,Cline:2012is}, which in particular can arise in many `mirror world'
models~\cite{Okun:2006eb,Foot:2014mia}.

\subsection{IV.2. Direct Detection}

We first neglect the impact of kinetic mixing between the twin and SM photons
on direct detection (DD) signatures and concentrate on the process of scattering purely
via Higgs exchange or by exchange of a twin scalar that mixes with the
Higgs. An interesting situation arises for $R \approx 1$. In this case,
$m_{\Delta '} \approx m_{\tau '}$ and therefore the Higgs couples to
both states with equal strength. On the other hand, the typical size of
the atom is
set by $a'_0 = (\alpha ' \mu_{\hat H})^{-1}$, which is $\approx 4 \ {\rm fm}$
for $\alpha ' \approx 10^{-2}$ and $m_{\hat H} \approx 20 \
\GeV$, values consistent with all constraints (see Figure~\ref{fig:atomR1}).
The size of the atomic system is thus
comparable to that of SM nuclei relevant for DD experiments, and the
possibility of a detectable `dark form factor' arises
(with the form factor approximately given by the Fourier transform of the 
ground state atomic wavefunction squared). 
While such a signal would be degenerate with modifications
to the DM halo velocity distribution for data from a single
DD experiment~\cite{Fox:2010bu}, multiple experiments with
different SM target nuclei could allow the dark form factor contribution
to be disentangled~\cite{Cherry:2014wia}.

Alternatively, if $R \gg 1$ then the atom's coupling
to the Higgs is dominantly through the $\Delta'$, whose
structure is on smaller scales than SM nuclei,
since $\Lambda_{\rm QCD}' > \Lambda_{\rm QCD}$. Thus,
in this case, we would have a basically momentum-independent dark
form factor, and spin independent cross sections
would be like those shown in Figure~\ref{fig:sigma}.

Finally, 
kinetic mixing between the two sectors can arise via
the operator $(\epsilon/2) F_{\mu \nu} {F'}^{\mu \nu}$.  This results in twin sector
particles acquiring SM-sector electric charges of size $\sim \epsilon e'$, with $e' =
\sqrt{4 \pi \alpha'}$. Low-energy radiative contributions to the kinetic mixing
parameter appear to be absent up to three-loop
order~\cite{TwinHiggs,FraternalTH}, and therefore one can expect $\epsilon \sim
(16 \pi^2)^{-4} \sim 10^{-9}$ \emph{if} a non-vanishing four-loop
contribution to $\epsilon$ indeed exists (UV contributions to kinetic
mixing can be present depending on the completion).   
Notice that our DM atoms are neutral under
both visible and twin sector electromagnetism and have vanishing
permanent electric dipole moments, due to their spherical distribution
of charge. Nevertheless, twin atoms have magnetic dipole moments
under both sectors, with the visible sector moment
suppressed by a factor of $\epsilon$. 
Experimental constraints on the size of $\epsilon$ arise from a combination of astrophysical, accelerator,
and direct detection considerations
\cite{Vogel:2013raa,Cline:2013zca,Kopp:2014tsa,Davidson:2000hf,Prinz:1998ua}.
The nature of the dominant constraint depends strongly on the values of $\alpha'$, $m_{\hat H}$ and $R$ chosen,
but for the range of parameters considered in this work, values of $\epsilon \lesssim 10^{-9}$ are likely to satisfy all current bounds.
 
\section{V. Conclusions}

We have shown that for the values of $\Lambda^\prime_{\rm QCD}$ allowed
by naturalness, and in the ungauged $U(1)'$ case, the twin hadron
$\Delta^\prime \sim b^\prime b^\prime b^\prime$
is a successful ADM candidate, with a mass, $\mathcal{O}(10 \GeV)$, automatically
in the most attractive regime for ADM theories 
to explain the $\mathcal{O}(1)$ ratio of DM-to-baryon energy densities.
If $U(1)'$ is gauged,  an asymmetric population of $\Delta '$
baryons is balanced by an equal number of asymmetric $\bar{\tau '}$. In
significant regions of parameter space, twin atoms are formed and are
successful DM candidates consistent with all current constraints, although
modified halo dynamics and direct detection signals are possible.

\begin{acknowledgments}

{\bf Acknowledgments:} 
We wish to thank N.~Craig, R.~Harnik, F.~Kahlhoefer, M.~McCullough \& the participants
of the CERN Neutral Naturalness workshop where this work was first presented for useful discussions.
IGG \& JMR are especially grateful to K.~Howe for many discussions of Twin Higgs theories.
RL and JMR thank, respectively, the Berkeley Center for Theoretical Physics, and the CERN Theory Group,
for hospitality during this work.  IGG \& RL acknowledge financial support from STFC studentships, and IGG
from a University of Oxford Scatcherd European Scholarship. 
\end{acknowledgments}


\bibliography{twinADM} 

\end{document}